\documentclass[12pt,a4paper]{article} 
\usepackage{amsmath}
\usepackage{amssymb}
\usepackage{latexsym}
\usepackage{amsthm}
\usepackage{amsbsy}
\usepackage{amstext}
\usepackage[dvips]{graphicx}

\def\lsim{\mathrel{\lower2.5pt\vbox{\lineskip=0pt\baselineskip=0pt 
           \hbox{$<$}\hbox{$\sim$}}}} 
\def\gsim{\mathrel{\lower2.5pt\vbox{\lineskip=0pt\baselineskip=0pt 
           \hbox{$>$}\hbox{$\sim$}}}}

\def\p{\partial}

\def\a{\alpha}

\def\e{\epsilon}
\def\da{\frac{\dot{a}}{a}}
\def\dda{\frac{\ddot{a}}{a}}
\def\db{\frac{\dot{b}}{b}}
\def\ddb{\frac{\ddot{b}}{b}}

\begin{document} 
\begin{flushright}
AUE-09-01\\
\end{flushright}

\vspace{10mm}

\begin{center}
{\Large \bf 
Anisotropic Evolution Driven by Kinetic Term}\\
\vspace{20mm}
 Masato Ito 
 \footnote{mito@auecc.aichi-edu.ac.jp}
\end{center}

\begin{center}
{
{}Department of Physics, Aichi University of Education, Kariya, 
448-8542, JAPAN
}
\end{center}

\vspace{25mm}

\begin{abstract}
 We present a simple model where anisotropic evolution is driven by
 kinetic term in extra dimensions. By introducing a canonical or a ghost
 kinetic term, the possibility of anisotropy is studied.
\end{abstract}
\newpage 
\baselineskip=6.0mm

If starting from theories such as string/M-theory or braneworld model,
it is believed that extra dimensions exist at
early universe
\cite{Gasperini:1992em,Quevedo:2002xw,Gasperini:2007vw,McAllister:2007bg}.
Since we live in $4$-dimensional spacetime, the extra
dimensions are invisible. 
In the string/M-theory, it is expected that extra
dimensions are compactified by moduli fields and some dynamics.
According to the picture of braneworld, we live in a 3-brane embedded in
higher dimensional world \cite{Randall:1999ee}. 
Moreover, the Randall-Sundrum model with warped metric have shown that
the fifth dimension is noncompact \cite{Randall:1999vf}.
However, whether the extra dimensions exist isn't confirmed by experiments till
now.

Recent precision of astronomical observations has illustrated whole history of
time evolution of our universe
\cite{Nolta:2008ih,Hinshaw:2008kr}.
Accelerated expansion of universe occurs at early time and late time,
the former is inflation and the latter is present cosmic acceleration
due to unknown dark energy.
One of the most significant problems of particle physics and cosmology is
to explain the whole history of our universe.
It is important to propose cosmological models to be consistent with
observations.
However it isn't easy to construct the consistent cosmological models
from the effective theory of string/M-theory.

Assuming that the birth of our universe is in higher dimensional spacetime, 
we suggest that the cosmological
model has anisotropic evolution of scale factors at early universe.
Accordingly $3$-dimensional space is expanding while extra dimensional
space is contracting.
In the framework of $4$-dimensional spacetime,
an example of vacuum cosmological model with anisotropy had been proposed by
Kasner \cite{kasner1,kasner2}.
The scale factor of each direction has anisotropic
power-law solution $t^{p_{i}}(i=1,2,3)$, and the
exponents are subject to constraints:
$\sum^{3}_{i=1}p_{i}=\sum^{3}_{i=1}p^{2}_{i}=1$.
The straightforward extension to higher
dimensional spacetime of the vacuum Kasner model have been
performed \cite{Chodos:1979vk,Levin:1994yw,Arkani-Hamed:1999gq}.
Several anisotropic models have been proposed
\cite{Cline:1999ky,Mazumdar:1999tk,Feinstein:2003iz,Chatterjee:2005sp,Greene:2007xu,
Koivisto:2008xf,Koivisto:2007bp}.
We would like to consider the anisotropic evolution in extra dimension,
far from the string/M-theory cosmologies.
In particular, we focus on the role of kinetic term.
For example it have been proposed that the kinetic term plays an important
role in cosmological models such as k-inflation
\cite{ArmendarizPicon:1999rj} or k-essence 
\cite{Chiba:1999ka,ArmendarizPicon:2000ah,Chiba:2002mw,Chimento:2003zf}. 

In this paper we present a simple model realizing
anisotropic evolution driven by kinetic term without potential.
We indicate that a canonical or a ghost kinetic term
leads to anisotropy.
The line element with homogeneous, anisotropic and spatially
flat is
\begin{align}
ds^{2}=-dt^{2}+a^{2}(t)\sum^{3}_{i=1}dx^{2}_{i}+b^{2}(t)\sum^{n}_{M=1}dy^{2}_{M}\,,
\label{eqn1}
\end{align}
where the scale factor of $3$-dimensional space is $a(t)$
and the scale factor of extra $n$-dimensional space is $b(t)$.
We consider a $(4+n)$-dimensional model with the kinetic term of a
homogeneous real scalar field $\phi(t)$ without potential.
The action is given by
\begin{align}
S=\int d^{4+n}x\;\sqrt{-g}\left(\frac{1}{2}{\cal R}
-\e\frac{1}{2}g^{\mu\nu}\p_{\mu}\phi\p_{\nu}\phi
\right)\,,
\label{eqn2}
\end{align}
where the higher dimensional gravitational constant is set to $1$.
By choosing the sign $\e=\pm 1$ of kinetic term,
we want to seek the possibility of anisotropic evolution.
The case of $\e=+1$ is canonical scalar field and the case of $\e=-1$ is
ghost field.
Likewise, anisotropic model with ghost kinetic term without
potential have been already proposed \cite{Patil:2005ii}.
The model is constructed under the assumptions that first time derivative for
Hubble parameter of each direction is zero and $\ddot{\phi}=0$ for ghost field.
In the model proposed here, the assumptions are relaxed.
The present model corresponds to the combination of both
\cite{Levin:1994yw} and \cite{Patil:2005ii}.

In the case of choice $\e=-1$, the existence of ghost field leads to the
instability of physical system due to unbounded kinetic energy.
Assuming that the birth of
our universe comes about by instability of physical system, 
the ghost field can be temporarily allowed. 
As argued in \cite{Patil:2005ii}, it is possible that the ghost field
generates from string theory, supergravity and
negative tension brane 
\cite{Kogan:2001qx,McFadden:2004se,Aref'eva:2004vw,Aref'eva:2004qr}.

From the setup of (\ref{eqn1}) and (\ref{eqn2}),
three components of the gravitational equation and equation of
motion for $\phi$ are given by
\begin{align}
&3\left(\da\right)^{2}+\frac{1}{2}n(n-1)\left(\db\right)^{2}+3n\da\db=
\e\frac{1}{2}\dot{\phi}^{2}
\,,\label{eqn3}\\
&2\dda+\left(\da\right)^{2}+n\ddb+\frac{1}{2}n(n-1)\left(\db\right)^{2}
 +2n\da\db=
-\e\frac{1}{2}\dot{\phi}^{2}
\,,\label{eqn4}\\
&3\dda+3\left(\da\right)^{2}
+(n-1)\ddb
+\frac{1}{2}(n-1)(n-2)\left(\db\right)^{2}
 +3(n-1)\da\db=
-\e\frac{1}{2}\dot{\phi}^{2}
\,,\label{eqn5}\\
&\e\left(\ddot{\phi}+\left(3\da+n\db\right)\dot{\phi}\right)=0
\,,\label{eqn6}
\end{align}
where the dot denotes the derivative with respect to $t$.
Here $\phi$ is properly normalized to be dimensionless.

We can seek the solutions to be consistent with
Eqs.(\ref{eqn3})-(\ref{eqn6}).
Consequently, two scale factors $a(t)$ and $b(t)$ becomes power-law
solutions or exponential-law solutions via the time-dependence of
$\phi(t)$.
Below we show two type of solutions separately.

{\bf Solution 1:}
The power-law solutions of scale factors are given by
\footnote{see Appendix}
\begin{align}
a(t)=a_{0}t^{p}\;,\;b(t)=b_{0}t^{q}\;,\;\phi(t)=\phi_{0}\log t+const\,,\label{eqn7}
\end{align}
where $a_{0},b_{0},\phi_{0}$ are real constants.
Moreover two exponents $p$ and $q$ are subject to the following constraints
\begin{align}
3p+nq=1\;,\;3p^{2}+nq^{2}=1-\e \phi^{2}_{0}\,.\label{eqn8}
\end{align}
When $\phi_{0}=0$, the model corresponds to vacuum Kasner model
\cite{Chodos:1979vk,Levin:1994yw,Arkani-Hamed:1999gq}.
Eq.(\ref{eqn8}) yields
\begin{align}
p=&\frac{3\pm\sqrt{3n\left(n+2-\e(n+3)\phi^{2}_{0}\right)}}{3(n+3)}\,,
\label{eqn9}\\
q=&\frac{n\mp\sqrt{3n\left(n+2-\e(n+3)\phi^{2}_{0}\right)}}{n(n+3)}\,.
\label{eqn10}
\end{align}
Each exponent depends on $n$, $\phi_{0}$ and $\e$.
We investigate the anisotropy for $\e=\pm 1$.

In the case of canonical scalar $(\e=+1)$, the reality of
exponents leads to
\begin{align}
0<\phi^{2}_{0}\leq \frac{n+2}{n+3}\,.\label{eqn11}
\end{align}
In particular, since square root is zero when
$\phi^{2}_{0}=(n+2)/(n+3)$, we obtain the isotropic evolution
corresponding to $p=q=1/(n+3)$.
Thus the anisotropy depends on the value of $0<\phi^{2}_{0}<(n+2)/(n+3)$.
Selecting the upper sign in Eqs.(\ref{eqn9}) and (\ref{eqn10}), 
$3$-dimensional space is decelerated expansion while extra dimensional space
is contracting.
The value of $\phi_{0}$ determines whether evolution is anisotropy or isotropy.

In the case of ghost $(\e=-1)$, the reality of
exponent is automatically satisfied.
For arbitrary $\phi_{0}$, the model has the phase of anisotropic
evolution $(p\neq q)$.
We consider that $3$-dimensional space is accelerated expansion. 
Since the condition is $p>1$ (power-law inflation) for upper sign, we have
\begin{align}
\phi^{2}_{0}>2\frac{n+2}{n}\,.\label{eqn12}
\end{align}
On the other hand, extra dimensional space becomes accelerated contraction.

Thus, for a canonical scalar (depending on $\phi_{0}$) or a ghost,
logarithmic dependence of $\phi(t)$ leads to the anisotropic power-law
solutions of scale factors.

{\bf Solution 2:}
The exponential-law solutions of scale factors are
given by\footnote{see Appendix}
\begin{align}
a(t)=a_{0}e^{\a t}\;,\;b(t)=b_{0}e^{-3\a t/n}\;,\;\phi(t)=\phi_{0}t+const\,,
\label{eqn13}
\end{align}
where $\phi_{0},\a$ are real constants
and are subject to
the following equation
\begin{align}
3(n+3)\a^{2}=-\e n\phi^{2}_{0}\,.\label{eqn14}
\end{align}
From Eq.(\ref{eqn14}), the case of canonical scalar $(\e=+1)$ is
excluded.
In the case of ghost $(\e=-1)$, we have
\begin{align}
\a=\phi_{0}\sqrt{\frac{n}{3(n+3)}}\,,\label{eqn15}
\end{align}
where sign $\pm$ due to square root is absorbed in $\phi_{0}$.

According to Eqs.(\ref{eqn13}) and (\ref{eqn15}), positive $\phi_{0}$
gives rise to accelerated expansion of our universe.
On the other hand, extra dimensional space is accelerated contraction.
Actually the solutions are described in \cite{Patil:2005ii}.
Thus, for a ghost, time dependence of $\ddot{\phi}(t)=0$ leads to the
anisotropic exponential-law solutions of scale factors.

In conclusion we have presented a simple model with anisotropy
driven by kinetic term in extra dimensions.
We suggest the possibility that the kinetic term plays an important role
at birth of our universe.
As shown in {\bf solution 1}, the canonical
or ghost kinetic term leads to the power-law anisotropy.
In the case of canonical kinetic term, $3$-dimensional space is
decelerated expansion and extra dimensions is accelerated contraction.
When the equality holds in Eq.(\ref{eqn11}), the
evolution becomes isotropy.
Thus the value of $\phi_{0}$ determines whether evolution is anisotropy
or isotropy.
In the case of ghost kinetic term, when inequality of Eq.(\ref{eqn12}) holds, 
$3$-dimensional space is
accelerated expansion (power-law inflation) and extra dimensions is
accelerated contraction.
The power-law anisotropy isn't discussed in \cite{Patil:2005ii}. 
As shown in {\bf solution 2}, ghost kinetic term leads to the
exponential-law anisotropy.
The case had been described in \cite{Patil:2005ii}.

In the case of allowing the temporarily ghost kinetic term,
it is assumed that the instability of dynamics triggers the birth of our universe.
We need some mechanisms in order that ghost kinetic energy is bounded.
In this paper we don't discuss it.

The model we considered here may be strange scenario.
The familiar early universe scenario is from inflaton-driven
inflation phase due to a nearly constant potential to reheating phase at
minimum of potential.
We suggest that the present model is applied to the birth of universe based
on anisotropy of higher dimension.
The mechanism of transition from anisotropic evolution to the familiar
inflation is needed.
Although the mechanism is unknown,
we emphasised an important role of kinetic term at early universe in the
paper.

{\bf Appendix:}
Solution 1 and 2 can be shown as follows.
Integrating (\ref{eqn6}), we obtain
\begin{align}
\dot{\phi}\propto a^{-3}b^{-n}\,.\label{eqn16}
\end{align}
The some combinations of Eqs.(\ref{eqn3})-(\ref{eqn5}) lead to the
following equations.
\begin{align}
(n+1)\times(\ref{eqn3})-3\times(\ref{eqn4})-n\times(\ref{eqn5})\to&\;
3\dda+n\ddb=-\e\dot{\phi}^{2}\,,\label{eqn17}\\
(\ref{eqn3})-(n-1)\times(\ref{eqn4})+n\times(\ref{eqn5})\to&\;
\dda+2\left(\da\right)^{2}+n\da\db=0\,,\label{eqn18}\\
(\ref{eqn3})+3\times(\ref{eqn4})-2\times(\ref{eqn5})\to&\;
\ddb+(n-1)\left(\db\right)^{2}+3\da\db=0\,.\label{eqn19}
\end{align}
Using Eqs.(\ref{eqn16})-(\ref{eqn19}),
it is easy to obtain the solutions.
In the case of $\dot{\phi}\propto t^{-1}$ ({\bf solution 1}), 
scale factors are power-law solution.
In the case of $\dot{\phi}\propto 1$ ({\bf solution 2}), 
scale factors are exponential-law solution.

{\bf Acknowledgements:}
M.I is financially supported by Aichi University of Education.


\end{document}